\documentclass[10pt,twocolumn,letterpaper]{article}
\usepackage[accsupp]{axessibility} 
\usepackage{iccv}
\usepackage{times}
\usepackage{graphicx}
\usepackage{amsmath}
\usepackage{amssymb}
\usepackage{commath}
\usepackage{multirow}
\usepackage{amsfonts,amsthm}
\usepackage{comment}
\usepackage[ruled,vlined]{algorithm2e}
\usepackage{xcolor}
\usepackage{bm, subfigure, url, pifont, overpic, cases}

\usepackage[pagebackref=true,breaklinks=true,letterpaper=true,colorlinks,bookmarks=false]{hyperref}

\iccvfinalcopy 

\def\iccvPaperID{5739} 

\ificcvfinal\fi

\begin{document}

\title{Towards Flexible Blind JPEG Artifacts Removal}

\author{Jiaxi Jiang\qquad\quad Kai Zhang$^{}$\thanks{Corresponding author.}\qquad\quad  Radu Timofte\vspace{0.1cm}\\
Computer Vision Lab, ETH Zurich, Switzerland\\
{\tt\small jiaxijiang@student.ethz.ch}\qquad
{\tt\small \{kai.zhang, timofter\}@vision.ee.ethz.ch}\\
\url{https://github.com/jiaxi-jiang/FBCNN}
}

\maketitle
\ificcvfinal\thispagestyle{empty}\fi

\begin{abstract}
Training a single deep blind model to handle different quality factors for JPEG image artifacts removal has been attracting considerable attention due to its convenience for practical usage. However, existing deep blind methods usually directly reconstruct the image without predicting the quality factor, thus lacking the flexibility to control the output as the non-blind methods. To remedy this problem, in this paper, we propose a flexible blind convolutional neural network, namely FBCNN, that can predict the adjustable quality factor to control the trade-off between artifacts removal and details preservation. Specifically, FBCNN decouples the quality factor from the JPEG image via a decoupler module and then embeds the predicted quality factor into the subsequent reconstructor module through a quality factor attention block for flexible control. Besides, we find existing methods are prone to fail on non-aligned double JPEG images even with only a one-pixel shift, and we thus propose a double JPEG degradation model to augment the training data. Extensive experiments on single JPEG images, more general double JPEG images, and real-world JPEG images demonstrate that our proposed FBCNN achieves favorable performance against state-of-the-art methods in terms of both quantitative metrics and visual quality.
\end{abstract}

\section{Introduction}
JPEG~\cite{wallace1992jpeg} is one of the most widely-used image compression algorithms and formats due to its simplicity and fast encoding/decoding speeds. JPEG compression splits an image into 8 $\times$ 8 blocks and applies discrete cosine transform (DCT)
to each block. The DCT coefficients are then divided by a quantization table and rounded to the nearest integer. The elements in the quantization table control the compression ratio and the rounding operation is the only lossy operation in the whole process. The quantization table is usually represented by an integer called quality factor (QF) ranging from 0 to 100, where a lower quality factor means less storage size but more lost information. Inspired by the success of deep neural networks (DNNs) for image classification~\cite{krizhevsky2012imagenet,vgg}, researchers began to resort to DNNs for JPEG artifacts removal and have achieved notable academic success.

However, existing methods for JPEG artifacts removal generally have four limitations in real applications: (1) Most existing DNNs based methods~\cite{cavigelli2017cas, chen2016trainable, dong2015compression, Liu_2018_CVPR_Workshops, zhang2019residual} trained a specific model for each quality factor, lacking the flexibility to learn a single model for different JPEG quality factors. (2) DCT based methods~\cite{ehrlich2020quantization, guo2016building, zhang2018dmcnn} need to obtain the DCT coefficients or quantization table as input, which is only stored in JPEG format. Besides, when images are compressed multiple times, only the most recent compression information is stored. (3) To solve the first problem, some recent work~\cite{ehrlich2020quantization, fu2019jpeg, zhang2017beyond} resort to training a single model for a large range of quality factors. However, these blind methods can only provide a deterministic reconstruction result for each input, ignoring the need for user preferences. (4) Existing methods are all trained with synthetic images which assumes that the low-quality images are compressed only once. However, most images from the Internet are compressed multiple times. Despite some progress for real recompressed images, \eg, from Twitter~\cite{dong2015compression, fu2019jpeg}, a detailed and complete study on double JPEG artifacts removal is still missing.

To tackle the above problems, we design a flexible blind convolutional neural network, namely FBCNN, for real JPEG image restoration. Our FBCNN is a single model that can deal with JPEG images with different quality factors. In addition, FBCNN can work independent of the image formats, as it directly processes images in pixel-domain, without the need to access the metadata of images. By further decoupling the latent quality factor from the input JPEG image, we can use this important parameter to guide the artifacts removal process. As a controllable variable with clear physical meaning, the predicted quality factor can also be adjusted via interactive selection to achieve a balance between artifacts removal and details preservation. To address the problems with real-world JPEG images, we provide a detailed study on the restoration of images with double JPEG compression. We find that existing blind methods are prone to fail when the 8 $\times$ 8 blocks of double JPEG compression are not aligned and $\text{QF}_1$ $\leq$ $\text{QF}_2$. However, our quality factor predictor can help to explain the behavior of current blind methods under unseen scenarios. We provide comprehensive empirical evidence showing that blind methods work are easy to be misled by the unseen compound artifacts, resulting in an unpleasant reconstructed output. By correcting the predicted quality factor, FBCNN instead can boost the performance on complex double JPEG images. To obtain a fully blind model, we further propose two solutions: correcting QF to the smaller one which can be estimated by our dominant QF estimation method or augmenting the training data with non-aligned double JPEG images.

To summarize, the main contributions of this paper are: 

(1) A flexible blind convolutional neural network for JPEG artifacts removal (FBCNN) is proposed. FBCNN can predict the latent quality factor to guide the image restoration. The predicted quality factor can be adjusted manually to control the preference between artifacts removal and details preservation according to the user's needs.

(2) We perform a thorough analysis of double JPEG images and provide solutions to take a step towards the restoration of real images. To the best of our knowledge, this is the first attempt to handle double non-aligned JPEG compression. We hope that the community will gradually begin to consider this more challenging and realistic scenario.

(3) We demonstrate the effectiveness of FBCNN on synthetic and real JPEG images with complex degradation settings. Our proposed FBCNN provides a useful solution for practical applications.

\section{Related Work}

\paragraph{JPEG Artifacts Removal Networks.} Learning-based methods have made notable progress in JPEG artifacts removal in the past few years. Dong~\etal~\cite{dong2015compression} first introduced deep learning to remove JPEG artifacts, inspired by the success of super-resolution network~\cite{dong2014learning}. Zhang~\etal~\cite{zhang2017beyond} employed batch normalization~\cite{ioffe2015batch} and residual learning~\cite{he2016deep} strategies to speed up the training process and boost the performance on general blind image restoration tasks. A wavelet transform based network was presented in~\cite{Liu_2018_CVPR_Workshops} as the generalization of dilated convolution~\cite{yu2015multi} and subsampling, leading to a large improvement. Fu~\etal~\cite{fu2019jpeg} proposed a deep convolutional sparse coding network that combines model-based methods with deep learning.
Besides, dual-domain convolutional network based methods~\cite{guo2016building, kim2020agarnet, zhang2018dmcnn, zheng2019implicit} were proposed to take advantage of redundancies on both pixel and DCT domains. Recently, Ehrlich~\etal~\cite{ehrlich2020quantization} trained their networks with the utilization of quantization table as prior information, which allows a single model to correct artifacts at any quality factor and achieved state-of-the-art results. 

\paragraph{Double JPEG Compression.} Double JPEG compression has been studied in the area of image forensics for a long time, as detection of double compression can provide important clues for the recovery of image processing history. Fu~\etal~\cite{fu2007generalized} showed that if an image has been JPEG compressed only once, then the first digits of the quantized JPEG coefficients follow a Benford-like logarithmic law. 
In~\cite{barni2010identification, bianchi2011analysis, chen2011detecting, luo2007novel}, double JPEG compression was classified into two cases: aligned and non-aligned. Chen~\etal~\cite{chen2011detecting} formulated the periodic characteristics of JPEG images in both spatial and DCT domains and showed that such periodic characteristics will be changed after recompression. Recently, learning-based methods~\cite{barni2017aligned, park2018double, wang2016double} were proposed to detect double JPEG compression. The estimation of the first quantization table of JPEG images is also a challenging problem and studied in both aligned~\cite{galvan2014first, pasquini2014multiple, xue2017mse, yu2016improved} and non-aligned cases~\cite{bianchi2012image, dalmia2018robust, yao2020improved}. However, these methods focus on analyzing the DCT coefficients, which are only stored in JPEG format. Besides, the research on double JPEG compression restoration is still missing.

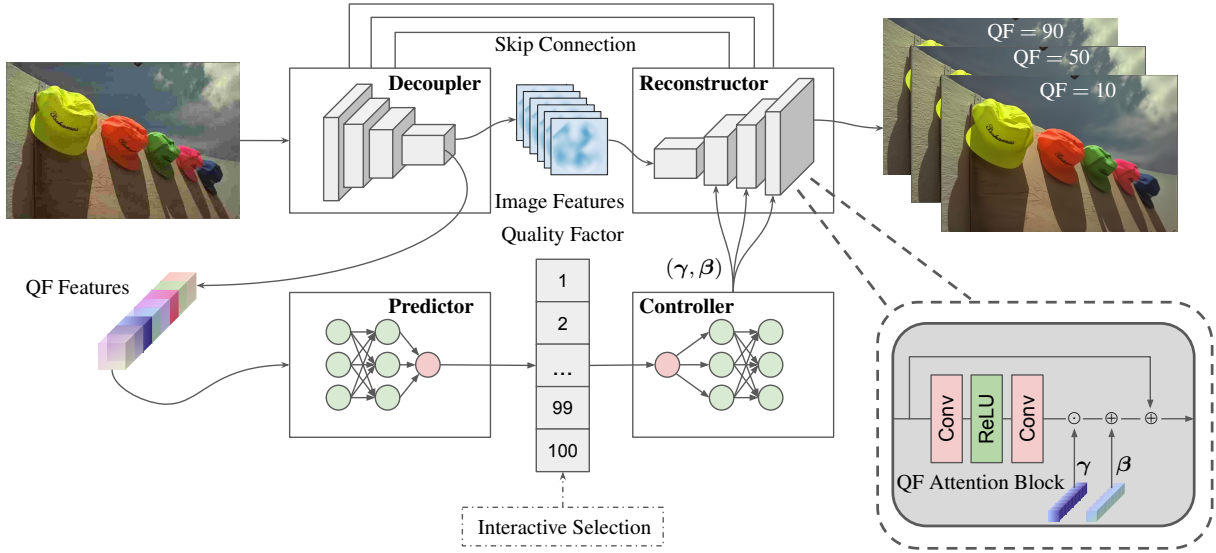
\begin{figure*}[htp]
\centering
\begin{overpic}[width=\textwidth]{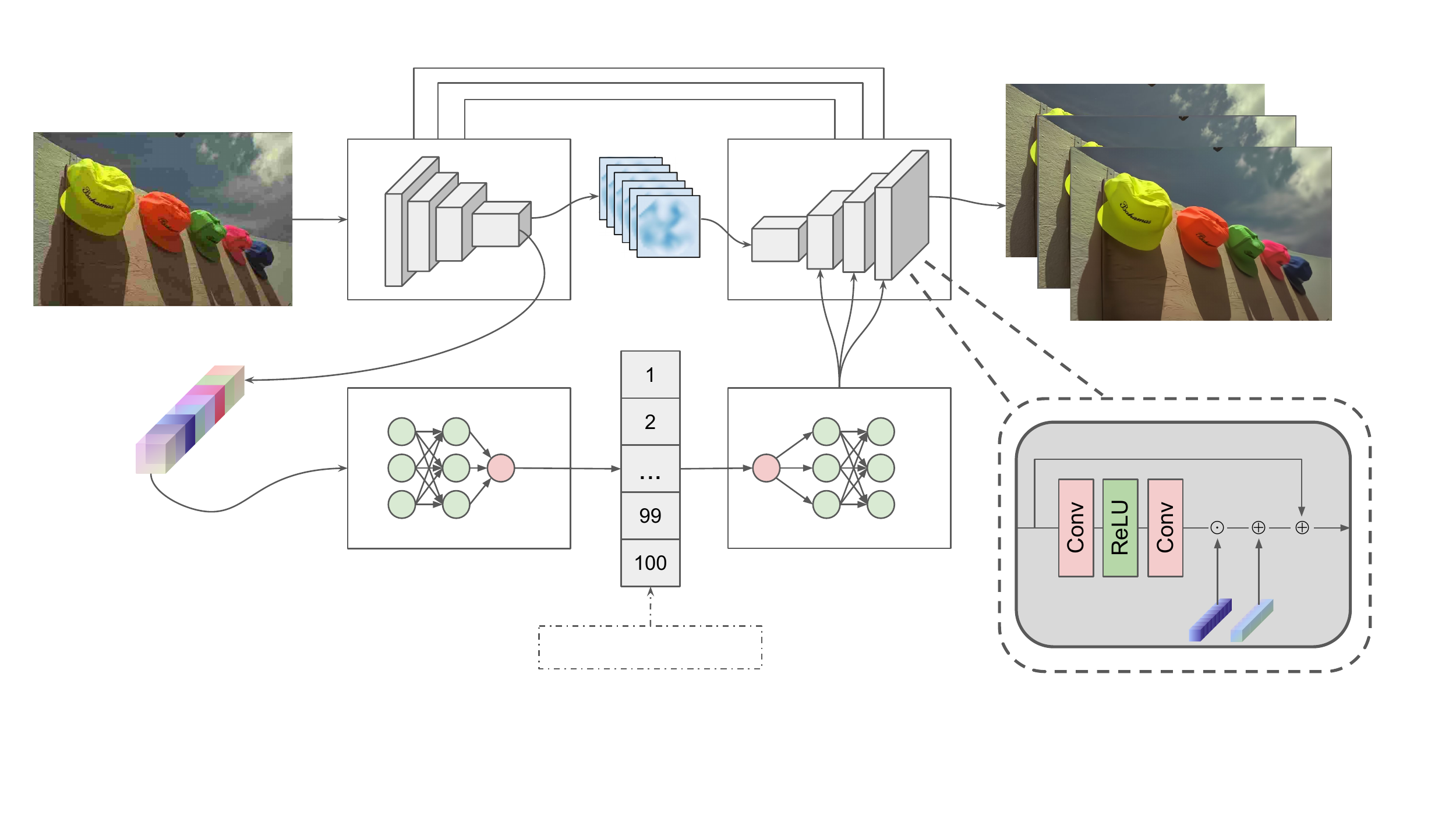}
\put(31.4,45){\color{black}{\footnotesize $\textbf{Decoupler}$}}
\put(50.5,45){\color{black}{\footnotesize $\textbf{Reconstructor}$}}
\put(31.4,28){\color{black}{\footnotesize $\textbf{Predictor}$}}
\put(50.5,28){\color{black}{\footnotesize $\textbf{Controller}$}}
\put(3.8,29.5){\color{black}{\footnotesize QF Features}}
\put(38.2,11.2){\color{black}{\footnotesize Interactive Selection}}
\put(70.1,14.6){\color{black}{\footnotesize QF Attention Block}}
\put(40,33.5){\color{black}{\footnotesize Quality Factor}}
\put(39.5,36){\color{black}{\footnotesize Image Features}}
\put(39.5,48){\color{black}{\footnotesize Skip Connection}}
\put(52.5,31){\color{black}{\footnotesize $(\boldsymbol{\gamma},\boldsymbol{\beta})$}}

\put(77,49){\color{white}{\footnotesize QF $=$ 90}}
\put(79,47){\color{white}{\footnotesize QF $=$ 50}}
\put(81,44.5){\color{white}{\footnotesize QF $=$ 10}}
\put(83.8,16){\color{black}{\footnotesize $\boldsymbol{\gamma}$}}
\put(86.8,16){\color{black}{\footnotesize $\boldsymbol{\beta}$}}

\end{overpic}
\vspace*{-17mm}
\caption{The architecture of the proposed FBCNN for JPEG artifacts removal. FBCNN consists of four parts, \ie, decoupler, quality factor predictor, flexible controller, and image reconstructor. The decoupler extracts the deep features from the input corrupted JPEG image and then splits them into image features and QF features which are subsequently fed into the reconstructor and predictor, respectively. The controller gets the estimated QF from the predictor and then generates QF embeddings. The QF attention block enables the controller to make the reconstructor produce different results according to different QF embeddings. The predicted quality factor can be changed with interactive selections to have a balance between artifacts removal and details preservation.}
\label{architecture}
\end{figure*}

\paragraph{Flexible Image Restoration.}
Flexible image generation based on the conditional variable has drawn much attention in \eg text-to-image generation \cite{li2019controllable, reed2016generative, xu2018attngan} and facial attribute editing~\cite{choi2018stargan, he2019attgan, liu2019stgan}.
However, these methods can not be directly adopted in image restoration. Zhang~\etal~\cite{zhang2018ffdnet} proposed to take a tunable noise level map as the input to handle noise on different levels. In~\cite{zhang2018learning}, a PCA-based dimensionality stretching of the degradation parameters was proposed to take blur kernel and noise level as input for super-resolution. Wang~\etal~\cite{wang2019cfsnet} proposed a novel controllable framework for interactive image restoration. He~\etal~\cite{he2020interactive} focused on the images with multiple degradations and added the multi-dimensional degradation information as input. These methods usually assume that the controllable variable is provided, but such information is almost unknown in real applications. This encourages us to work towards a flexible blind solution for image restoration.

\section{Proposed Method}
In this section, we first introduce the architecture of our FBCNN, and then present its advantage over other state-of-the-art methods, especially for practical recompressed JPEG images.

\subsection{Flexible Blind Artifacts Removal Network}
Fig.~\ref{architecture} illustrates the overall architecture of our proposed FBCNN. FBCNN is an end-to-end model which takes a JPEG image as input and directly generates the output image. Specifically, FBCNN consists of four components: decoupler, QF predictor, flexible controller, and image reconstructor. The network is fairly straightforward, with each component designed to achieve a specific task.

\paragraph{Decoupler:} The decoupler aims to extract the deep features and decouple the latent quality factor from the input image. It involves four scales, each of which has an identity skip connection to the reconstructor. 4 residual blocks are adopted in each scale, and each residual block is composed of two 3 $\times$ 3 convolution layers with ReLU activation in the middle. 2 $\times$ 2 strided convolutions are adopted for the downscaling operations. The number of output channels in each layer from the first to the fourth scale is set to 64, 128, 256, 512, respectively. The image features from the decoupler are passed into the reconstructor. At the same time, they are also shared by an additional quality factor branch that uses residual blocks to extract higher-level information, followed by a global average pooling layer to get the global quality factor features from the image features.

\paragraph{Quality Factor Predictor:} The QF predictor is a 3-layer MLP (multilayer perceptron) that takes as input the 512-dimensional QF features and produces an estimated quality factor $\text{QF}_\text{est}$ of the compressed image. We set the number of nodes in each hidden layer as 512 for a better prediction. During training, patches with small sizes may only include limited information and correspond to multiple quality factors so that the quality factor can not be accurately estimated, which may lead to an unstable training process. Therefore, we use the L1 loss function to avoid too much penalty for such outliers. Let $N$ be the batch size during training, the loss for quality factor estimation in each batch can be written as:
\begin{equation}
\mathcal{L}_\text{QF} = \frac{1}{N}\sum_{i=1}^{N} \norm{\text{QF}_\text{est}^i - \text{QF}_\text{gt}^i}_1.
\end{equation}

\paragraph{Flexible Controller:} The flexible controller is a 4-layer MLP and takes as input the quality factor, representing the degree of compression of the targeted image. The controller aims to learn an embedding of the given quality factor that can be fused into the reconstructor for flexible control. Inspired by recent research in spatial feature transform~\cite{park2019semantic, wang2018recovering}, the controller learns a mapping function that outputs a modulation parameter pair $(\boldsymbol{\gamma}, \boldsymbol{\beta})$ which embeds the given quality factor. Specifically, the first three layers of MLP generate shared intermediate conditions, which are then split into three parts corresponding to the three scales in the reconstructor. In the last layer of MLP, we learn different parameter pairs for different scales in reconstructor whereas shared $(\boldsymbol{\gamma}, \boldsymbol{\beta})$ are broadcasted to the QF Attention block within the same scale.

\paragraph{Image Reconstructor:} The image reconstructor includes three scales and receives image features from decoupler and quality factor embedding parameters~$(\boldsymbol{\gamma}, \boldsymbol{\beta})$ to generate the restored clean image. The QF attention block is an important component of the reconstructor. The number of QF attention blocks in each scale is set to 4. The learned parameter pair~$(\boldsymbol{\gamma}, \boldsymbol{\beta})$ adaptively influences the outputs by applying an affine transformation spatially to each intermediate feature map inside the QF attention block of each scale.

After obtaining~$(\boldsymbol{\gamma}, \boldsymbol{\beta})$~from the controller, the transformation is carried out by scaling and shifting feature maps of a specific layer:
\begin{equation}
    \boldsymbol{F_\text{out}} = \boldsymbol{\gamma}\odot \boldsymbol{F_\text{in}}\oplus \boldsymbol{\beta},
\end{equation}
where $\boldsymbol{F_\text{in}}$ and $\boldsymbol{F_\text{out}}$ denote the feature maps before and after the affine transformation, and $\odot$ is referred to as element-wise multiplication.

Given $N$ training samples within a batch, the goal of the image reconstructor is to minimize the following L1 loss function between reconstructed image $\mathbf{I_\text{rec}}$ and the original ground-truth image $\mathbf{I_\text{gt}}$:
\begin{equation}
	\mathcal{L}_\text{rec} = \frac{1}{N}\sum_{i=1}^{N}\norm{\mathbf{I}_\text{rec}^i - \mathbf{I}_\text{gt}^i}_1.
\end{equation}
Overall, the complete training objective can be written as:
\begin{equation}
	\mathcal{L}_\text{total} = \mathcal{L}_\text{rec} + \lambda\cdot\mathcal{L}_\text{QF},
\end{equation}
where $\lambda$ controls the balance between image reconstruction and QF estimation.

\subsection{Comparison with Other Design Choices}
In the following, we will clarify the differences between the proposed FBCNN and two alternative design choices.

\paragraph{A blind model without QF prediction:} Existing blind methods only provide a deterministic result, ignoring the need of the user's preference. Besides, as we will discuss in Sec.~\ref{sec: flexiblity}, although the pure blind model performs favorably for single JPEG artifacts removal without knowing the quality factor, it does not generalize well to real corrupted images whose artifacts are more complex. FBCNN can be viewed as multiple deblockers and can control the trade-off between JPEG artifacts removal and details preservation.

\paragraph{Cascaded QF prediction and non-blind model:} It is also possible to design a QF predictor cascaded by a non-blind method like CBD-Net~\cite{guo2019toward}. However, our method enjoys some benefits compared with such a cascaded design: First, for accurate quality factor estimation, a convolutional network starting from the same scale as the input image is needed, which would increase the total model size and cost more training and inference time. Instead, we only add a relatively small prediction branch. Second, our decoupler shared parameters for QF estimation and image reconstruction, accelerating the convergence of predicting QF. On the contrary, in cascaded design, inaccurate QF estimation would lead to an unstable training process. It might be a solution to train a QF predictor and then freeze it to train the second part for reconstruction. Nevertheless, it would cost more training time than our joint training schedule. Fourth, in cascaded networks, the predicted parameter is treated as the input of the second part and propagates through the whole encoder-decoder architecture. Instead, our predicted parameter QF is the only input to the decoder part. We can change the QF to adjust different outputs during inference without the need to change the encoded image features, which saves half of the inference time.

\begin{figure*}[t!]
\centering
\hspace{9mm}
\begin{overpic}[trim=0cm 0cm 0cm 0cm,clip=true,width=0.15\linewidth]{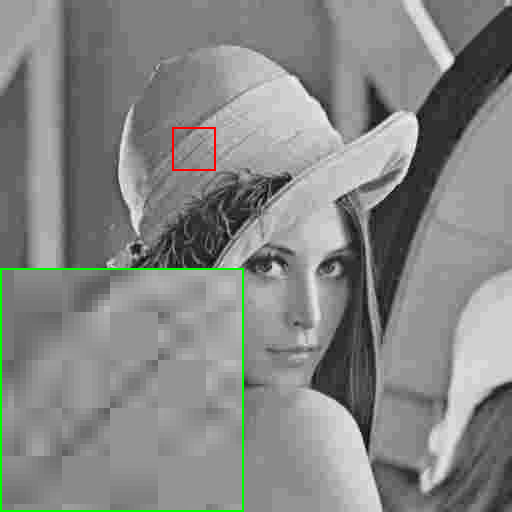}
\put(35,105){\color{black}{QF=10}}
\put(-38,45){\color{black}{JPEG}}
\end{overpic}
\begin{overpic}[trim=0cm 0cm 0cm 0cm,clip=true,width=0.15\linewidth]{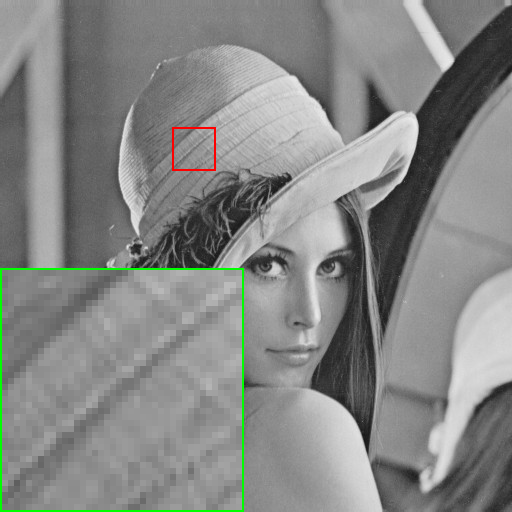}
\put(32,105){\color{black}{QF=90}}
\end{overpic}
\begin{overpic}[trim=0cm 0cm 0cm 0cm,clip=true,width=0.15\linewidth]{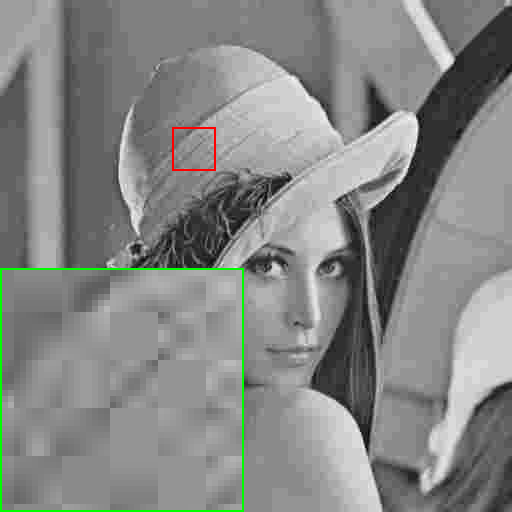}
\put(18,105){\color{black}{QF=(90,10)}}
\end{overpic}
\begin{overpic}[trim=0cm 0cm 0cm 0cm,clip=true,width=0.15\linewidth]{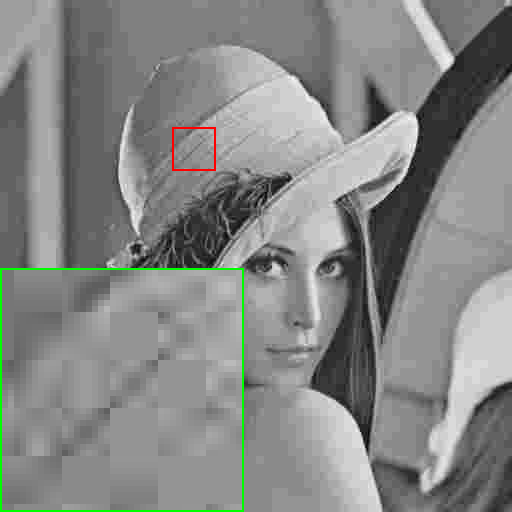}
\put(18,105){\color{black}{QF=(10,90)}}
\end{overpic}
\begin{overpic}[trim=0cm 0cm 0cm 0cm,clip=true,width=0.15\linewidth]{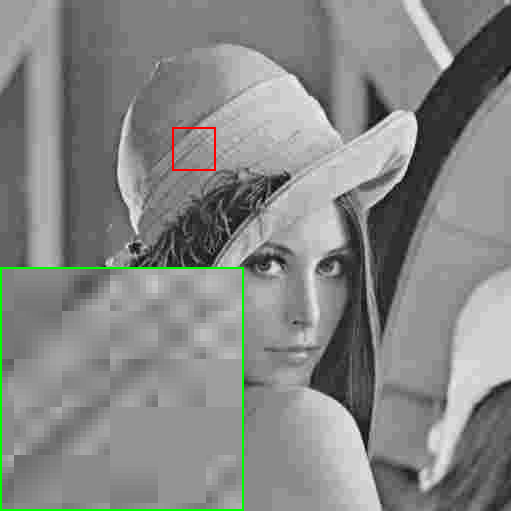}
\put(15,105){\color{black}{QF=(90,10)*}}
\end{overpic}
\begin{overpic}[trim=0cm 0cm 0cm 0cm,clip=true,width=0.15\linewidth]{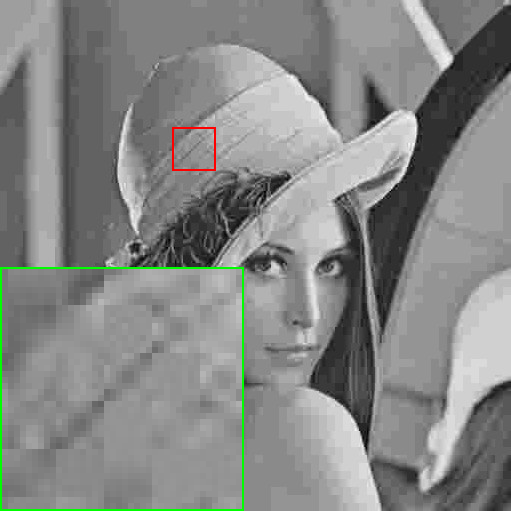}
\put(15,105){\color{black}{QF=(10,90)*}}
\end{overpic}\\
\vspace{0.6mm}
\hspace{9mm}
\begin{overpic}[trim=0cm 0cm 0cm 0cm,clip=true,width=0.15\linewidth]{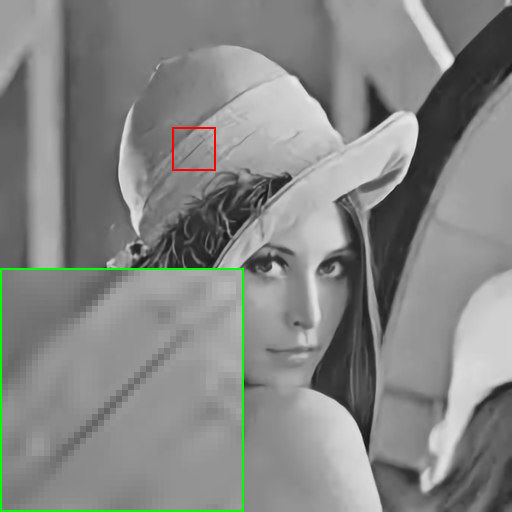}
\put(-46,45){\color{black}{DnCNN}}
\end{overpic}
\begin{overpic}[trim=0cm 0cm 0cm 0cm,clip=true,width=0.15\linewidth]{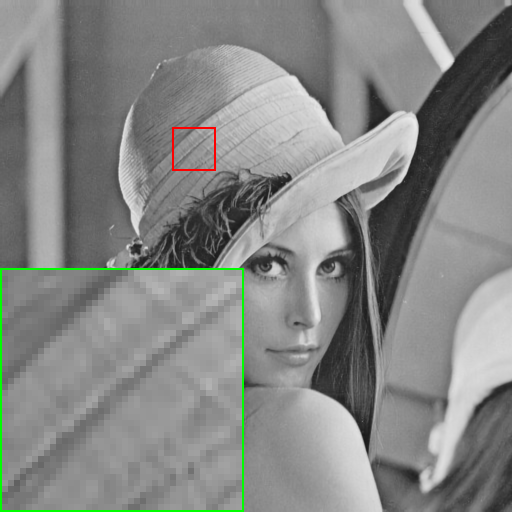}
\put(40,85){\color{black}{\footnotesize }}
\end{overpic}
\begin{overpic}[trim=0cm 0cm 0cm 0cm,clip=true,width=0.15\linewidth]{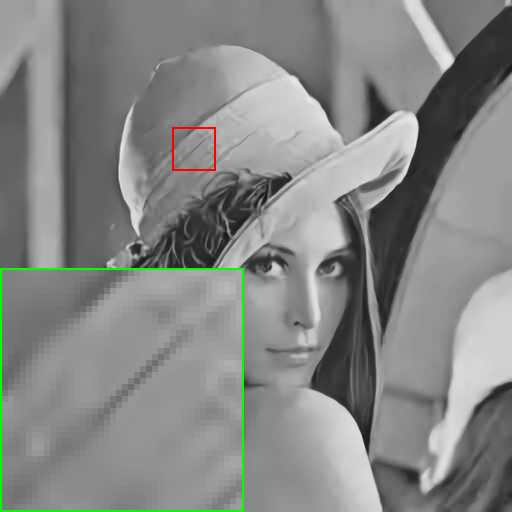}
\put(40,85){\color{black}{\footnotesize }}
\end{overpic}
\begin{overpic}[trim=0cm 0cm 0cm 0cm,clip=true,width=0.15\linewidth]{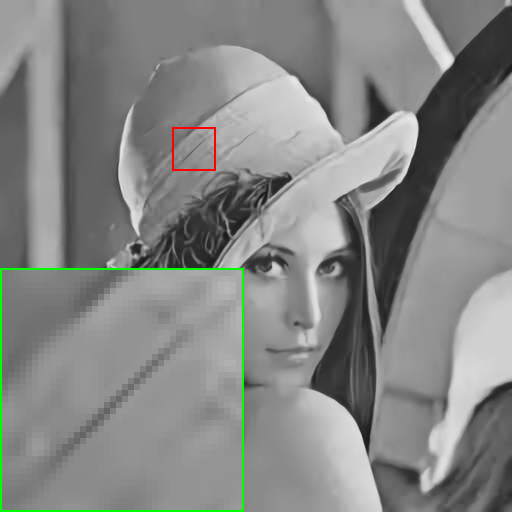}
\put(40,85){\color{black}{\footnotesize }}
\end{overpic}
\begin{overpic}[trim=0cm 0cm 0cm 0cm,clip=true,width=0.15\linewidth]{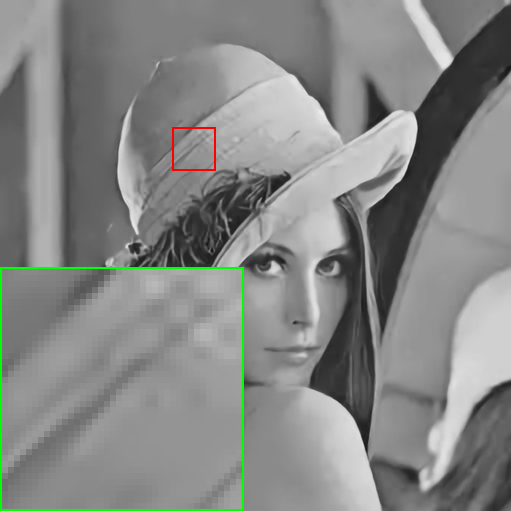}
\put(40,85){\color{black}{\footnotesize }}
\end{overpic}
\begin{overpic}[trim=0cm 0cm 0cm 0cm,clip=true,width=0.15\linewidth]{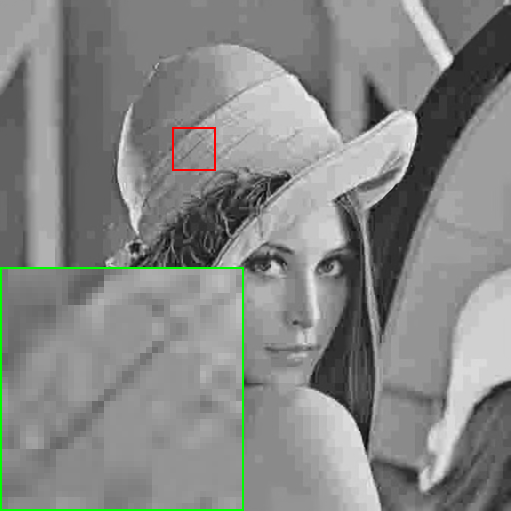}
\put(40,85){\color{black}{\footnotesize }}
\end{overpic}\\
\vspace{0.6mm}
\hspace{9mm}
\begin{overpic}[trim=0cm 0cm 0cm 0cm,clip=true,width=0.15\linewidth]{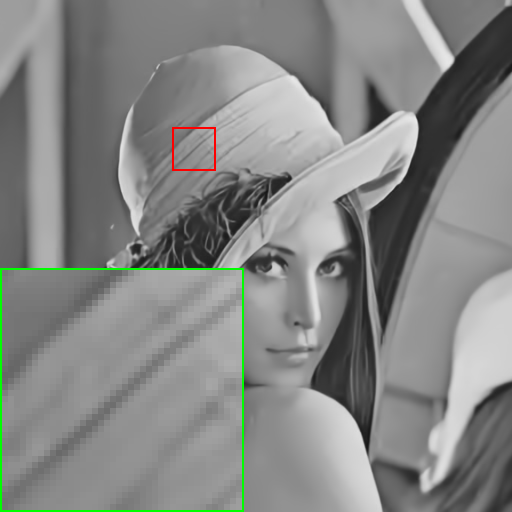}
\put(-43,45){\color{black}{QGAC}}
\end{overpic}
\begin{overpic}[trim=0cm 0cm 0cm 0cm,clip=true,width=0.15\linewidth]{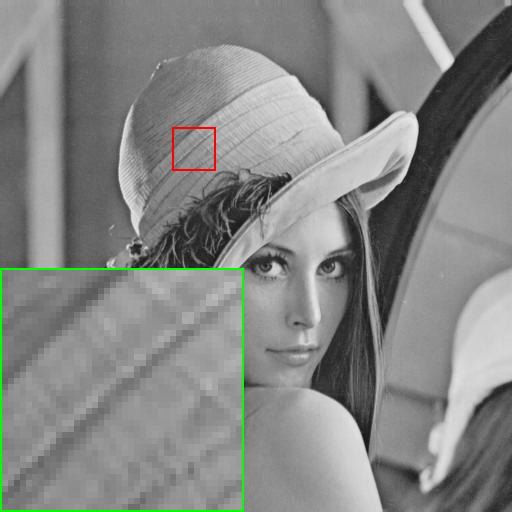}
\put(40,85){\color{black}{\footnotesize }}
\end{overpic}
\begin{overpic}[trim=0cm 0cm 0cm 0cm,clip=true,width=0.15\linewidth]{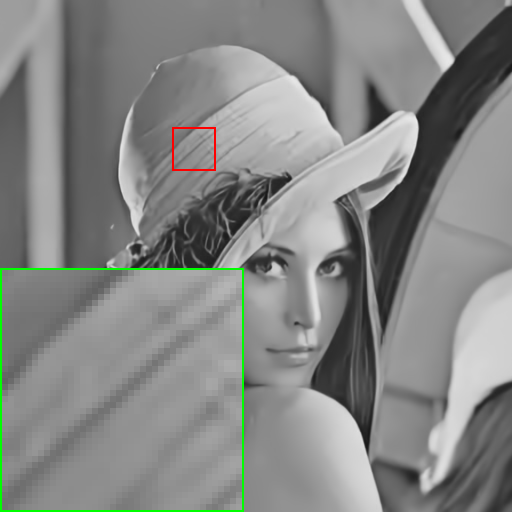}
\put(40,85){\color{black}{\footnotesize }}
\end{overpic}
\begin{overpic}[trim=0cm 0cm 0cm 0cm,clip=true,width=0.15\linewidth]{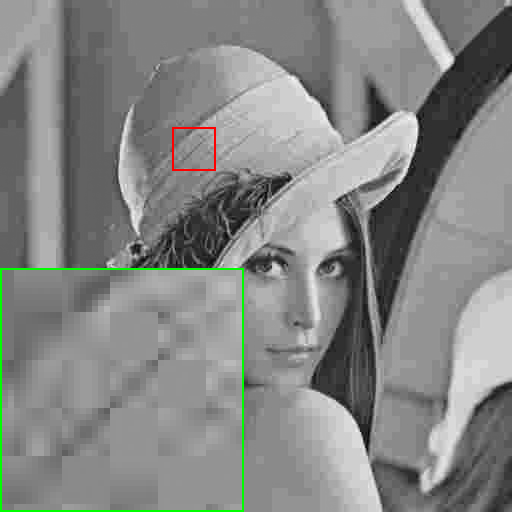}
\put(40,85){\color{black}{\footnotesize }}
\end{overpic}
\begin{overpic}[trim=0cm 0cm 0cm 0cm,clip=true,width=0.15\linewidth]{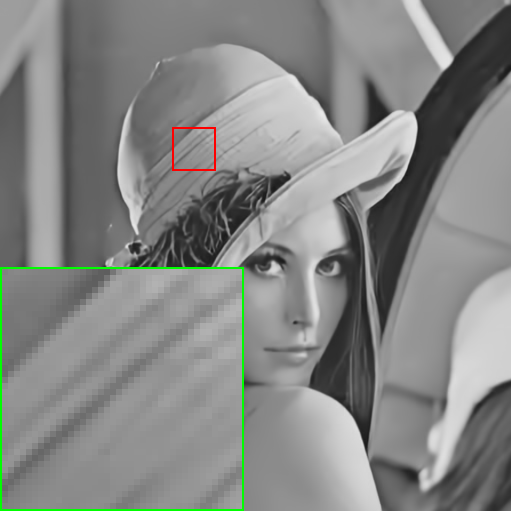}
\put(40,85){\color{black}{\footnotesize }}
\end{overpic}
\begin{overpic}[trim=0cm 0cm 0cm 0cm,clip=true,width=0.15\linewidth]{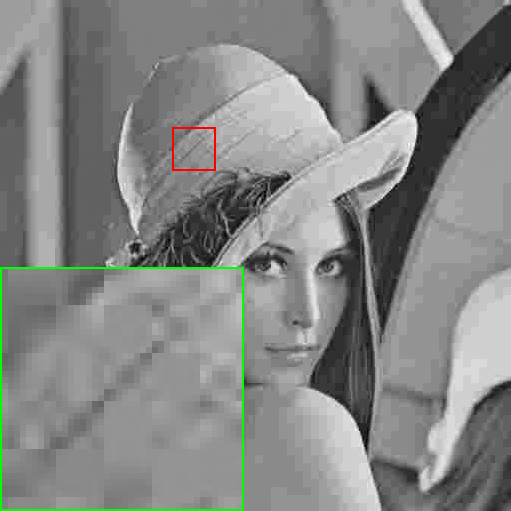}
\put(40,85){\color{black}{\footnotesize }}
\end{overpic}
\caption{Visual comparisons of a JPEG image with different degradation settings and their restored results by DnCNN and QGAC. QF $=$ ($\text{QF}_1$, $\text{QF}_2$) denotes that the image is firstly compressed with $\text{QF}_1$ and then compressed with $\text{QF}_2$. `*' means there is a pixel shift (1,1) between blocks of two compression. Even only a shift of one pixel between two compression can lead to failures of existing methods.}
\label{fig: djpeg_analysis}
\end{figure*}

\subsection{Restoration of Double JPEG Images}
\label{sec: flexiblity}
\paragraph{Limitations of existing methods:} Although some existing work claimed to work on recompressed JPEG images, a detailed study on the restoration of double JPEG compression is still missing. We find that the current blind methods always fail when the blocks of two JPEG compression are not aligned and $\text{QF}_1$ $\leq$ $\text{QF}_2$, even if there is an only one-pixel shift between two compression.

Let us look at an example in Fig.~\ref{fig: djpeg_analysis}, where the appearances of JPEG images with different compression settings can be observed. To get non-aligned double JPEG images, we remove the first row and the first column of the image between the first compression with $\text{QF}_1$ and the second one with $\text{QF}_2$. For aligned double JPEG with QF $=$ (90, 10), (10, 90), and non-aligned double JPEG with QF $=$ (90, 10)*, the blocking effects are similar to single compression with QF $=$ 10: the edges of 8 $\times$ 8 blocks are apparent. However, in the case of non-aligned double JPEG with QF $=$ (10, 90)*, the blocking edges are not clear anymore. We test representative blind methods DnCNN~\cite{zhang2017beyond} and QGAC~\cite{ehrlich2020quantization} on these images. 

As shown in Fig.~\ref{fig: djpeg_analysis}, in cases of QF $=$ 90, 10, (90, 10), the blocking effects are well removed by both methods. DnCNN also works well on QF $=$ (10, 90), while QGAC fails in this case because QGAC extracts the quantization table from the JPEG image, but JPEG images only keep the most recent compression information. Therefore, we conclude that existing quantization table-based methods are not suitable for real application.

However, in the case of non-aligned double JPEG compression when $\text{QF}_1$ $=$ 10 and $\text{QF}_2$ $=$ 90, both methods do not work. Since our FBCNN is also a pixel-based blind method like DnCNN but can predict the quality factor, it can be used to explain the behavior behind a blind method. We test FBCNN using the same images. Not surprisingly, we get a similar, almost unchanged reconstructed result, but we find the predicted quality factor is 90. We continue to test other images with non-aligned double JPEG compression and $\text{QF}_1$ $<$ $\text{QF}_2$, finding that the predicted quality factor is always close to $\text{QF}_2$. This is to say, blind methods trained with single JPEG compression image pairs are always misled by the appearance of non-aligned double JPEG images with $\text{QF}_1$ $<$ $\text{QF}_2$. They also do not work when $\text{QF}_1$ $=$ $\text{QF}_2$.

In summary, we classify double JPEG compression into two categories: simple and complex compression. Simple compression corresponds to non-aligned double JPEG with $\text{QF}_1$ $>$ $\text{QF}_2$ and all aligned double JPEG compression, which is actually equivalent to single JPEG compression. Complex compression corresponds to non-aligned double JPEG with $\text{QF}_1$ $\leq$ $\text{QF}_2$, where composite artifacts occur. We test images with these degradation settings by a recent double JPEG compression algorithm~\cite{park2018double}, finding that only images with non-aligned double JPEG with $\text{QF}_1$ $\leq$ $\text{QF}_2$ can be identified as double JPEG compression, which further support our arguments.

To overcome the problem with non-aligned double JPEG compression, we propose two solutions, from the perspectives of adjusting the QF to utilize our flexible network and augmenting the training data.

\begin{table*}[hbt!]
\caption{PSNR$\vert$SSIM$\vert$PSNRB results of different methods on \textbf{grayscale} JPEG images with \textbf{single} compression. Please note that the methods marked with * train a specific model for each quality factor. The best two results are highlighted in 
\vspace*{-4mm}
\textcolor{red}{red} and \textcolor{blue}{blue} colors, respectively.}
\begin{center}

\resizebox{\linewidth}{!}{

\begin{tabular}{ c c |c c c c c c c } 
\hline
{Dataset} & Quality &{JPEG}  & {ARCNN*} & {MWCNN*}  & {DnCNN} & {DCSC} & {QGAC} & {FBCNN~(Ours)}\\
\hline
 \multirow{4}{4em}{Classic5}
 &10 & 27.82$\vert$0.760$\vert$25.21 & 29.03 $\vert$0.793$\vert$28.76 & \textcolor{blue}{30.01}$\vert$\textcolor{blue}{0.820}$\vert$\textcolor{blue}{29.59}&  29.40$\vert$0.803$\vert$29.13& 29.62$\vert$0.810$\vert$29.30 & 29.84$\vert$0.812$\vert$29.43  &  \textcolor{red}{30.12}$\vert$\textcolor{red}{0.822}$\vert$\textcolor{red}{29.80} \\ 
 &20& 30.12$\vert$0.834$\vert$27.50 & 31.15$\vert$0.852$\vert$30.59 &\textcolor{blue}{32.16}$\vert$\textcolor{blue}{0.870}$\vert$\textcolor{blue}{31.52}& 31.63$\vert$0.861$\vert$31.19&  31.81$\vert$0.864$\vert$31.34 & 31.98$\vert$0.869$\vert$31.37 & \textcolor{red}{32.31}$\vert$\textcolor{red}{0.872}$\vert$\textcolor{red}{31.74} \\ 
 &30 & 31.48$\vert$0.867$\vert$28.94 & 32.51$\vert$0.881$\vert$31.98 &\textcolor{blue}{33.43}$\vert$\textcolor{blue}{0.893}$\vert$\textcolor{blue}{32.62}& 32.91$\vert$0.886$\vert$32.38&  33.06$\vert$0.888$\vert$32.49 & 33.22$\vert$0.892$\vert$32.42 & \textcolor{red}{33.54}$\vert$\textcolor{red}{0.894}$\vert$\textcolor{red}{32.78} \\ 
 &40 & 32.43$\vert$0.885$\vert$29.92 & 33.32$\vert$0.895$\vert$32.79 &\textcolor{blue}{34.27}$\vert$\textcolor{blue}{0.906}$\vert$\textcolor{blue}{33.35}& 33.77$\vert$0.900$\vert$33.23&  33.87$\vert$0.902$\vert$33.30 & 34.05$\vert$0.905$\vert$33.12 & \textcolor{red}{34.35}$\vert$\textcolor{red}{0.907}$\vert$\textcolor{red}{33.48} \\ 

\hline

 \multirow{4}{4em}{LIVE1} 
 &10 & 27.77$\vert$0.773$\vert$25.33 & 28.96$\vert$0.808$\vert$28.68 &\textcolor{blue}{29.69}$\vert$\textcolor{blue}{0.825}$\vert$\textcolor{blue}{29.32}& 29.19$\vert$0.812$\vert$28.90&  29.34$\vert$0.818$\vert$29.01 &29.51$\vert$\textcolor{blue}{0.825}$\vert$29.13 & \textcolor{red}{29.75}$\vert$\textcolor{red}{0.827}$\vert$\textcolor{red}{29.40} \\ 
 &20& 30.07$\vert$0.851$\vert$27.57 & 31.29$\vert$0.873$\vert$30.76 &\textcolor{blue}{32.04}$\vert$\textcolor{red}{0.889}$\vert$\textcolor{blue}{31.51}& 31.59$\vert$0.880$\vert$31.07&  31.70$\vert$0.883$\vert$31.18 & 31.83$\vert$\textcolor{blue}{0.888}$\vert$31.25 & \textcolor{red}{32.13}$\vert$\textcolor{red}{0.889}$\vert$\textcolor{red}{31.57} \\ 
 &30 & 31.41$\vert$0.885$\vert$28.92 & 32.67$\vert$0.904$\vert$32.14 &\textcolor{blue}{33.45}$\vert$\textcolor{blue}{0.915}$\vert$\textcolor{blue}{32.80}& 32.98$\vert$0.909$\vert$32.34& 33.07$\vert$0.911$\vert$32.43 & 33.20$\vert$0.914$\vert$32.47 & \textcolor{red}{33.54}$\vert$\textcolor{red}{0.916}$\vert$\textcolor{red}{32.83} \\ 
 &40 & 32.35$\vert$0.904$\vert$29.96 & 33.61$\vert$0.920$\vert$33.11 &\textcolor{blue}{34.45}$\vert$\textcolor{blue}{0.930}$\vert$\textcolor{red}{33.78}& 33.96$\vert$0.925$\vert$33.28&  34.02$\vert$0.926$\vert$33.36 & 34.16$\vert$0.929$\vert$33.36 & \textcolor{red}{34.53}$\vert$\textcolor{red}{0.931}$\vert$\textcolor{blue}{33.74} \\ 

\hline
 \multirow{4}{4em}{BSDS500} 
 &10 & 27.80$\vert$0.768$\vert$25.10  & 29.10$\vert$0.804$\vert$28.73 & \textcolor{blue}{29.61}$\vert$\textcolor{blue}{0.820}$\vert$\textcolor{blue}{29.14}& 29.21$\vert$0.809$\vert$28.80& 29.32$\vert$0.813$\vert$28.91 & 29.46$\vert$\textcolor{red}{0.821}$\vert$28.97 & \textcolor{red}{29.67}$\vert$\textcolor{red}{0.821}$\vert$\textcolor{red}{29.22} \\ 
 &20& 30.05$\vert$0.849$\vert$27.22 & 31.28$\vert$0.870$\vert$30.55& \textcolor{blue}{31.92}$\vert$\textcolor{red}{0.885}$\vert$\textcolor{blue}{31.15}& 31.53$\vert$0.878$\vert$30.79&  31.63$\vert$0.880$\vert$30.92 & 31.73$\vert$\textcolor{blue}{0.884}$\vert$30.93 & \textcolor{red}{32.00}$\vert$\textcolor{red}{0.885}$\vert$\textcolor{red}{31.19} \\ 
 &30 & 31.37$\vert$0.884$\vert$28.53 & 32.67$\vert$0.902$\vert$31.94& \textcolor{blue}{33.30}$\vert$\textcolor{blue}{0.912}$\vert$\textcolor{red}{32.34}& 32.90$\vert$0.907$\vert$31.97&  32.99$\vert$0.908$\vert$32.08 & 33.07$\vert$\textcolor{blue}{0.912}$\vert$32.04 & \textcolor{red}{33.37}$\vert$\textcolor{red}{0.913}$\vert$\textcolor{blue}{32.32} \\ 
 &40 & 32.30$\vert$0.903$\vert$29.49 & 33.55$\vert$0.918$\vert$32.78& \textcolor{blue}{34.27}$\vert$\textcolor{red}{0.928}$\vert$\textcolor{red}{33.19}& 33.85$\vert$0.923$\vert$32.80&  33.92$\vert$0.924$\vert$32.92 & 34.01$\vert$\textcolor{blue}{0.927}$\vert$32.81 & \textcolor{red}{34.33}$\vert$\textcolor{red}{0.928}$\vert$\textcolor{blue}{33.10} \\ 
 
 \hline
\end{tabular}
}
\end{center}
\label{tab: singlecompare}
\end{table*}

\paragraph{FBCNN trained with a single JPEG degradation model with dominant QF correction:} Since our FBCNN can provide different outputs by setting different quality factors, correcting the predicted QF to the smaller one, which actually dominates the main compression, is expected to improve the restoration results. However, to get a fully blind model, it is crucial to infer the smaller quality factor automatically. By utilizing the property of JPEG compression, we find that the quality factor of a JPEG image with single compression can be obtained by doing another JPEG compression with all possible QFs. The image's QF corresponds to the global minimum of the MSE (mean squared error) between two JPEG images. We further extend this method to challenging non-aligned double JPEG images with $\text{QF}_1$ $<$ $\text{QF}_2$. We apply another JPEG compression with all possible QFs after a shift in the range of 0 to 7 in two directions. We also calculate the MSE curves for each shift possibility between the two JPEG images. For each MSE curve, we search for the first minimum. It can be found that among all the first minimums, the QF at the smallest first minimum is always close to $\text{QF}_1$, while the QF at the global minimum is approximate to $\text{QF}_2$. Besides, we constrain the MSE of the smallest first minimum to be smaller than a threshold $T$ to have more robust results. We empirically set $T$ to 30 in our experiment. We name the FBCNN model with dominant QF correction as FBCNN-D.

\vspace{-0.2cm}
\paragraph{FBCNN trained with double JPEG degradation model:} We can also solve this problem by augmenting the training data using images with double JPEG compression. We propose a new degradation model to synthesize the non-aligned double JPEG image $\mathbf{y}$ from the uncompressed image $\mathbf{x}$ via
\begin{equation}
\mathbf{y}=\text{JPEG}(\text{shift}(\text{JPEG}(\mathbf{x}, \text{QF}_1)), \text{QF}_2).
\end{equation}
For shift operation, we randomly remove the first $i$ rows and $j$ columns of the image after the first compression, where $0\leq i,j\leq7$. When trained with double JPEG compressed images, the weight of quality factor loss is set to zero. Then the dominant quality factor can be trained in an unsupervised way. We name the FBCNN model with augmented training data as FBCNN-A. Note that our double JPEG degradation model can also be applied to other tasks such as blind single image super-resolution~\cite{zhang2021designing}.

\begin{figure*}[hbt]
\centering
\subfigure[JPEG  (29.64dB)]{
\includegraphics[trim=0cm 1cm 0cm 0cm,clip=true,width=0.23\linewidth]{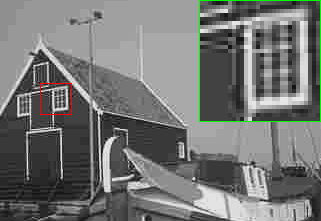}
}
\subfigure[ARCNN (31.15dB)]{
\includegraphics[trim=0cm 1cm 0cm 0cm,clip=true,width=0.23\linewidth]{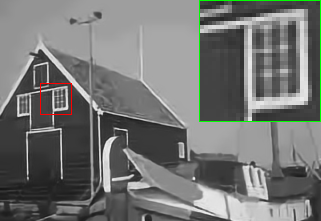}
}
\subfigure[MWCNN (32.38dB)]{
\includegraphics[trim=0cm 1cm 0cm 0cm,clip=true,width=0.23\linewidth]{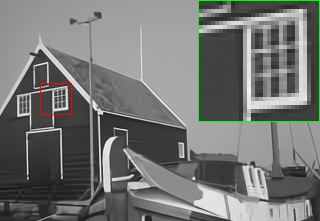}
}
\subfigure[DnCNN (31.36dB)]{
\includegraphics[trim=0cm 1cm 0cm 0cm,clip=true,width=0.23\linewidth]{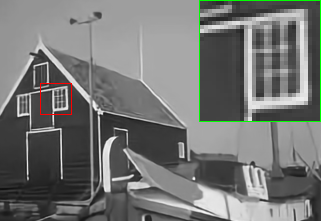}
}
\subfigure[DCSC (31.68dB)]{
\includegraphics[trim=0cm 1cm 0cm 0cm,clip=true,width=0.23\linewidth]{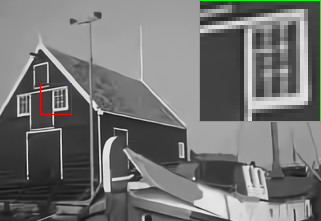}
}
\subfigure[QGAC (31.97dB)]{
\includegraphics[trim=0cm 1cm 0cm 0cm,clip=true,width=0.23\linewidth]{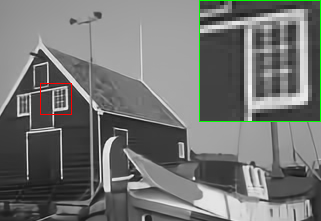}
}
\subfigure[FBCNN (32.51dB)]{
\includegraphics[trim=0cm 1cm 0cm 0cm,clip=true,width=0.23\linewidth]{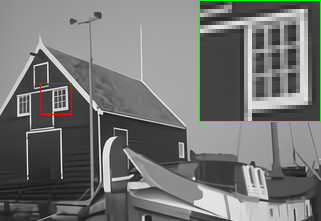}
}
\subfigure[Ground Truth]{
\includegraphics[trim=0cm 1cm 0cm 0cm,clip=true,width=0.23\linewidth]{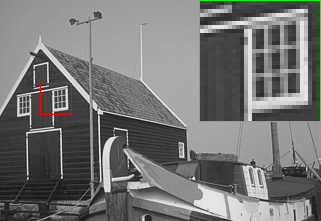}
}
\caption{Visual comparisons of different methods on a \textbf{single} JPEG image `BSDS500: 140088' with QF $=$ 10.}
\label{fig: singlecompare}
\end{figure*}

\section{Experiments}
\subsection{Data Preparation and Network Training}
For fair comparisons, JPEG images used during training and evaluation are all generated by the MATLAB JPEG encoder. We use the Y channel of YCbCr space for grayscale image comparison, and the RGB channels for color image comparison. Following \cite{ehrlich2020quantization}, we employ DIV2K \cite{agustsson2017ntire} and Flickr2K \cite{timofte2017ntire} as our training data. During training, we randomly extract patch pairs with the size 128 $\times$ 128, and the quality factor is randomly sampled from 10 to 95. We set $\lambda$ to 0.1. To optimize the parameters of FBCNN, we adopt the Adam solver \cite{kingma2014adam} with batch size 256. The learning rate starts from $1 \times 10^{-4}$ and decays by a factor of $0.5$ every $4 \times 10^4$ iterations and finally ends with $1.25 \times 10^{-5}$. We train our model with PyTorch on eight NVIDIA GeForce GTX 2080Ti GPUs. It takes about two days to obtain FBCNN.

\subsection{Single JPEG Image Restoration}
\paragraph{Grayscale JPEG image restoration} We first evaluate the performance of the proposed FBCNN on images with single JPEG compression. We test on the commonly used benchmarks: Classic5~\cite{zeyde2010single}, LIVE1 \cite{sheikh2005live} and the test set of BSDS500~\cite{martin2001database}. We compare our proposed FBCNN with ARCNN~\cite{dong2015compression}, MWCNN~\cite{Liu_2018_CVPR_Workshops}, DnCNN~\cite{zhang2017beyond}, DCSC~\cite{fu2019jpeg}, QGAC~\cite{ehrlich2020quantization}. It should be pointed out that ARCNN and MWCNN train a single network for each specific value of quality factor, and DCSC is trained with quality factors from 10 to 40. Only DnCNN, QGAC, and our FBCNN cover a full range of quality factors. We calculate the PSNR, SSIM, and PSNR-B for quantitative assessment. The quantitative results are shown in Table \ref{tab: singlecompare}. Our method has significantly better results than other blind methods and moderately better results than MWCNN, which trains each model for a specific quality factor. For subjective comparisons, some restored images of different approaches on the LIVE1 dataset have been presented. As can be seen in Fig.~\ref{fig: singlecompare}, the results of our FBCNN are more visually pleasing.

\begin{table}[!b]
\begin{center}
\caption{PSNR$\vert$SSIM$\vert$PSNRB results of QGAC and FBCNN-C on \textbf{color} JPEG images with \textbf{single} compression.}
\label{tablecolor}
\resizebox{\linewidth}{!}{
\begin{tabular}{ c c |c c c} 
\hline
Dataset & QF &JPEG  & QGAC & FBCNN-C~(Ours) \\

\hline
 \multirow{4}{4em}{LIVE1}
 &10 & 25.69$\vert$0.743$\vert$24.20 
 & 27.62$\vert$0.804$\vert$27.43 
 &27.77$\vert$0.803$\vert$27.51\\
 
 &20 & 28.06$\vert$0.826$\vert$26.49
 & 29.88$\vert$0.868$\vert$29.56
 &30.11$\vert$0.868$\vert$29.70\\
 
  &30 & 29.37$\vert$0.861$\vert$27.84 
 & 31.17$\vert$0.896$\vert$30.77 
 &31.43$\vert$0.897$\vert$30.92\\
 
  &40 & 30.28$\vert$0.882$\vert$28.84
 & 32.05$\vert$0.912$\vert$31.61 
 &32.34$\vert$0.913$\vert$31.80\\
 
\hline
 \multirow{4}{4em}{BSDS500}
 &10 & 25.84$\vert$0.741$\vert$24.13 
 & 27.74$\vert$0.802$\vert$27.47 
 &27.85$\vert$0.799$\vert$27.52\\
 
 &20 & 28.21$\vert$0.827$\vert$26.37 
 & 30.01$\vert$0.869$\vert$29.53 
 &30.14$\vert$0.867$\vert$29.56\\
 
  &30 & 29.57$\vert$0.865$\vert$27.72 
 & 31.33$\vert$0.898$\vert$30.70
 &31.45$\vert$0.897$\vert$30.72\\
 
  &40 & 30.52$\vert$0.887$\vert$28.69 
 & 32.25$\vert$0.915$\vert$31.50
 &32.36$\vert$0.913$\vert$31.52\\
 \hline
  \multirow{4}{4em}{ICB}
 &10 & 29.44$\vert$0.757$\vert$28.53 
 & 32.06$\vert$0.816$\vert$32.04 
 &32.18$\vert$0.815$\vert$32.15\\
 
 &20 & 32.01$\vert$0.806$\vert$31.11 
 & 34.13$\vert$0.843$\vert$34.10 
 &34.38$\vert$0.844$\vert$34.34\\
 
  &30 & 33.20$\vert$0.831$\vert$32.35 
 & 35.07$\vert$0.857$\vert$35.02
 &35.41$\vert$0.857$\vert$35.35\\
 
  &40 & 33.95$\vert$0.840$\vert$33.14 
 & 32.25$\vert$0.915$\vert$31.50&36.02$\vert$0.866$\vert$35.95\\
 \hline
\end{tabular}
}
\end{center}
\end{table}

\begin{figure*}[hbt]
\centering
\subfigure[JPEG (27.45dB)]{
\includegraphics[trim=0cm 0cm 0cm 0cm,clip=true,width=0.23\linewidth]{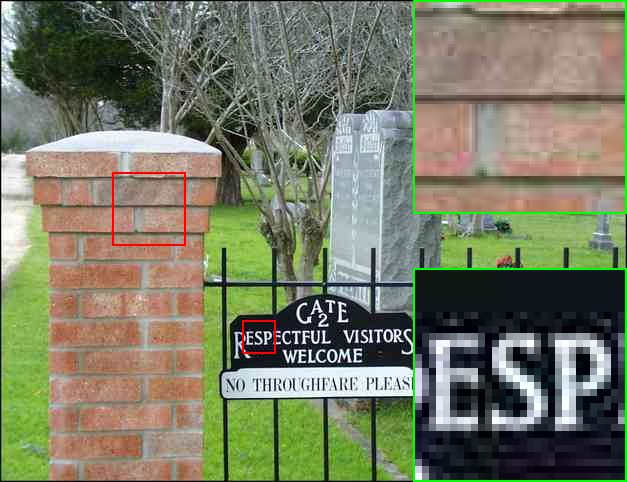}
}
\subfigure[QF = 10 (28.13dB)]{
\includegraphics[trim=0cm 0cm 0cm 0cm,clip=true,width=0.23\linewidth]{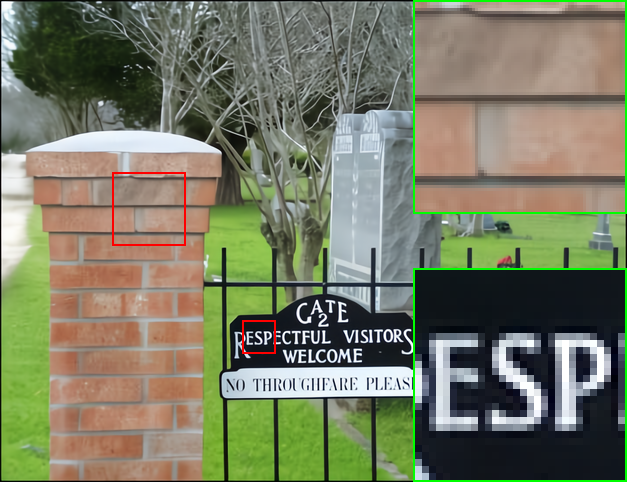}
}
\subfigure[QF = 30 (29.34dB)]{
\includegraphics[trim=0cm 0cm 0cm 0cm,clip=true,width=0.23\linewidth]{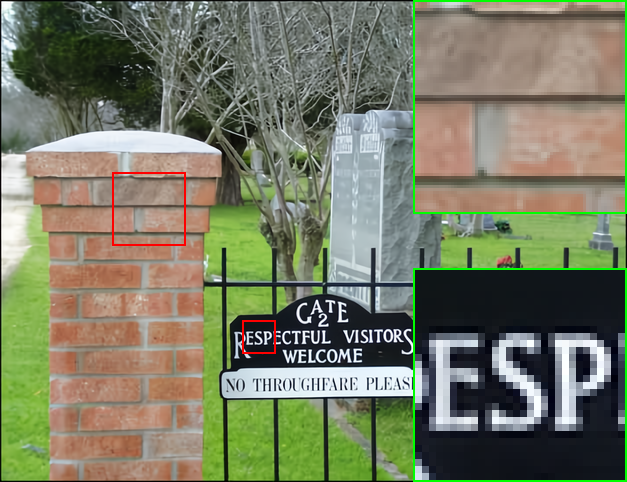}
}
\subfigure[QF = 90 (28.05dB)]{
\includegraphics[trim=0cm 0cm 0cm 0cm,clip=true,width=0.23\linewidth]{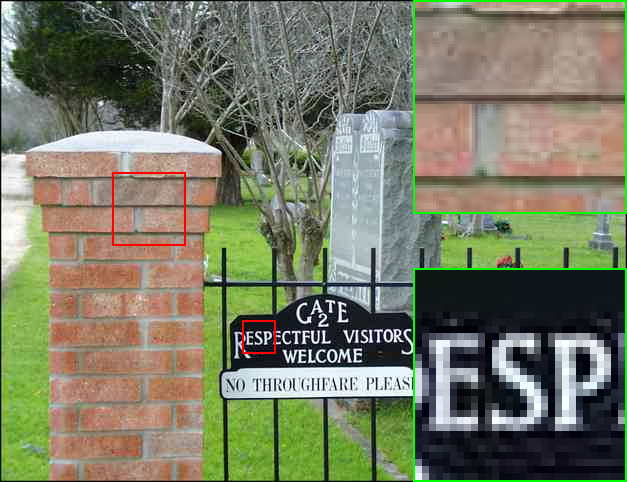}
}
\caption{An example to show the flexibility of FBCNN by setting different QFs into the network. The JEPG image is `LIVE1: cemetry' compressed with quality factor 30. Although the artifacts around the words can be effectively removed when the set QF is small, the texture on the bricks becomes blurred. Users can get the desired results according to their preference through interactive selection by FBCNN.}
\label{fig: flexibility}
\end{figure*}

\begin{table*}[h!]
\caption{PSNR$\vert$SSIM$\vert$PSNRB results of different methods on \textbf{grayscale} JPEG images with \textbf{non-aligned} \textbf{double} compression. The testing images are synthesized from the LIVE1 dataset. The best two results are highlighted in \textcolor{red}{red} and \textcolor{blue}{blue} colors, respectively.} 
\vspace*{-4mm}
\begin{center}
\resizebox{\linewidth}{!}{
\begin{tabular}{ c c| c c c c c c c} 
\hline
Type & QF & {JPEG} & {DnCNN} & {DCSC} & {QGAC}  & {FBCNN (Ours)} & {FBCNN-D (Ours)} & {FBCNN-A (Ours)}\\
\hline
 \multirow{3}{4em}{$\text{QF}_1$$>$$\text{QF}_2$} 
 &(30,10)& 27.49$\vert$0.762$\vert$25.62 & 
 28.95$\vert$0.805$\vert$28.61&  
 29.08$\vert$0.810$\vert$28.81 & 
 29.24$\vert$\textcolor{blue}{0.818}$\vert$28.94 & 
 \textcolor{red}{29.46}$\vert$\textcolor{red}{0.820}$\vert$\textcolor{blue}{29.11}& 
\textcolor{red}{29.46}$\vert$\textcolor{red}{0.820}$\vert$29.10 &
\textcolor{blue}{29.44}$\vert$\textcolor{blue}{0.818}$\vert$\textcolor{red}{29.12}\\ 
 
 &(50,10)  & 27.65$\vert$0.769$\vert$25.69 & 29.13$\vert$0.810$\vert$28.76&  
29.25$\vert$0.815$\vert$28.96 & 29.42$\vert$\textcolor{blue}{0.823}$\vert$29.08 & 
\textcolor{blue}{29.64}$\vert$\textcolor{red}{0.825}$\vert$\textcolor{red}{29.23}&
\textcolor{red}{29.65}$\vert$\textcolor{red}{0.825}$\vert$\textcolor{blue}{29.22} &
29.61$\vert$\textcolor{blue}{0.823}$\vert$29.20\\ 
 
 &(50,30) & 30.62$\vert$0.866$\vert$28.85 & 32.20$\vert$0.895$\vert$31.50&  32.30$\vert$0.897$\vert$31.78 & 32.32$\vert$0.899$\vert$31.72 & 
 \textcolor{blue}{32.61}$\vert$\textcolor{red}{0.902}$\vert$31.88& 
 \textcolor{blue}{32.61}$\vert$\textcolor{red}{0.902}$\vert$\textcolor{blue}{31.89} &
 \textcolor{red}{32.69}$\vert$\textcolor{blue}{0.901}$\vert$\textcolor{red}{32.24}
 \\ 

\hline

 \multirow{3}{4em}{$\text{QF}_1$$=$$\text{QF}_2$} 
 &(10,10)& 26.48$\vert$0.715$\vert$25.08 & 27.73$\vert$0.765$\vert$27.49&  
 27.76$\vert$0.768$\vert$27.59 & 27.78$\vert$0.771$\vert$27.59 & 
 \textcolor{blue}{27.96}$\vert$\textcolor{blue}{0.774}$\vert$\textcolor{blue}{27.75}& 27.95$\vert$\textcolor{blue}{0.774}$\vert$27.74&
 \textcolor{red}{28.25}$\vert$\textcolor{red}{0.777}$\vert$\textcolor{red}{28.14} \\ 
 
 &(30,30)  & 29.98$\vert$0.847$\vert$28.53 & 31.40$\vert$0.878$\vert$30.86&  31.48$\vert$0.880$\vert$31.10 & 31.43$\vert$0.881$\vert$30.99 & 
 31.64$\vert$\textcolor{blue}{0.884}$\vert$\textcolor{blue}{31.14}&
\textcolor{blue}{31.65}$\vert$\textcolor{blue}{0.884}$\vert$\textcolor{blue}{31.14}&
 \textcolor{red}{31.94}$\vert$\textcolor{red}{0.886}$\vert$\textcolor{red}{31.73}\\ 
 
 &(50,50)  & 31.58$\vert$0.888$\vert$30.18 & 33.12$\vert$0.912$\vert$32.44&  33.28$\vert$0.914$\vert$32.80 & 33.12$\vert$0.914$\vert$32.50 & 
33.38$\vert$\textcolor{blue}{0.917}$\vert$32.61&
\textcolor{blue}{33.45}$\vert$0.914$\vert$\textcolor{blue}{32.85}&
 \textcolor{red}{33.70}$\vert$\textcolor{red}{0.919}$\vert$\textcolor{red}{33.34} \\ 

 \hline

 \multirow{3}{4em}{$\text{QF}_1$$<$$\text{QF}_2$} \ 
 &(10,30) & 27.55$\vert$0.760$\vert$26.94 & 28.33$\vert$0.790$\vert$28.17&  28.31$\vert$0.789$\vert$28.19 & 28.30$\vert$0.791$\vert$28.18 & 
 28.29$\vert$0.791$\vert$28.15& 
 \textcolor{blue}{28.94}$\vert$\textcolor{blue}{0.802}$\vert$\textcolor{blue}{28.82} &\textcolor{red}{29.38}$\vert$\textcolor{red}{0.816}$\vert$\textcolor{red}{29.30}\\ 
 
 &(10,50) & 27.69$\vert$0.768$\vert$27.41 & 28.30$\vert$0.791$\vert$28.24&  28.40$\vert$0.794$\vert$28.35 & 28.23$\vert$0.791$\vert$28.18 & 
 28.20$\vert$0.789$\vert$28.14& \textcolor{blue}{28.96}$\vert$\textcolor{blue}{0.801}$\vert$\textcolor{blue}{28.88} &\textcolor{red}{29.52}$\vert$\textcolor{red}{0.820}$\vert$\textcolor{red}{29.45}\\ 
 
 &(30,50)  & 30.61$\vert$0.865$\vert$29.60 & 31.89$\vert$0.890$\vert$31.46&  32.08$\vert$0.893$\vert$31.78 & 31.81$\vert$0.891$\vert$31.43 & 31.96$\vert$0.893$\vert$31.50& 
 \textcolor{blue}{32.31}$\vert$\textcolor{blue}{0.895}$\vert$\textcolor{blue}{31.94}&
 \textcolor{red}{32.64}$\vert$\textcolor{red}{0.900}$\vert$\textcolor{red}{32.49} \\ 
 
\hline

\end{tabular}
}
\end{center}
\label{doublejpegtable}
\end{table*}

\begin{figure*}[h!]
\centering
\subfigure[JPEG (31.34dB)]{
\includegraphics[trim=0cm 3cm 0cm 0cm,clip=true,width=0.23\linewidth]{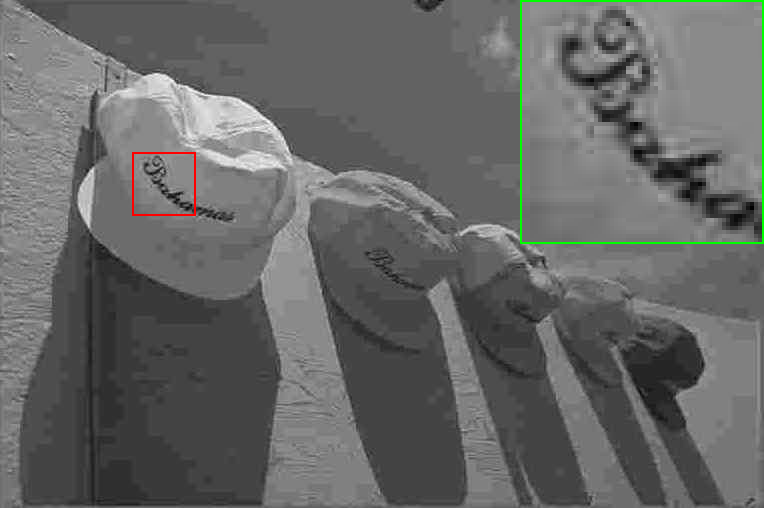}}
\subfigure[DnCNN (32.10dB)]{
\includegraphics[trim=0cm 3cm 0cm 0cm,clip=true,width=0.23\linewidth]{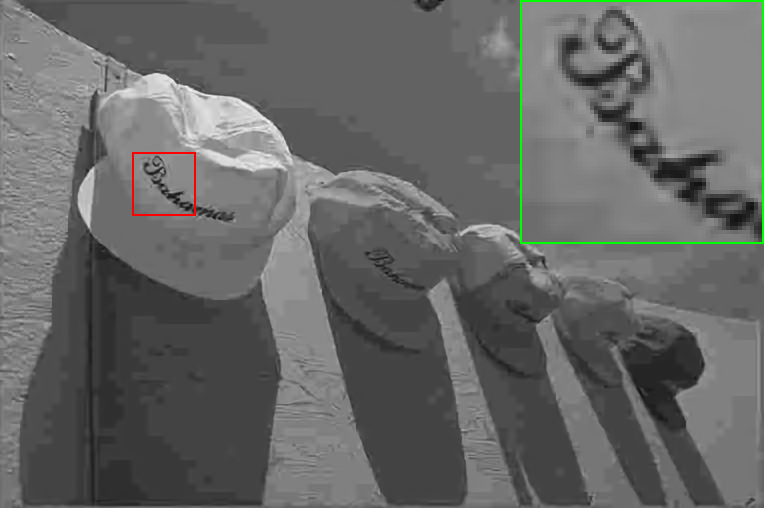}}
\subfigure[DCSC (31.97dB)]{
\includegraphics[trim=0cm 3cm 0cm 0cm,clip=true,width=0.23\linewidth]{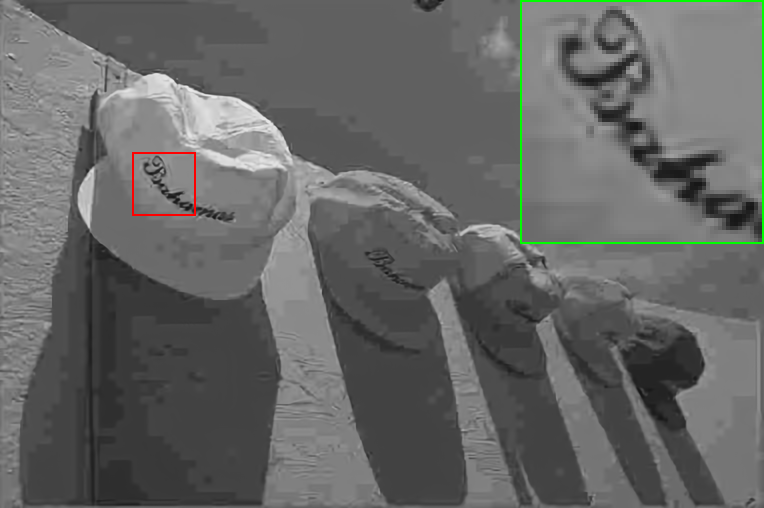}}
\subfigure[QGAC (32.06dB)]{
\includegraphics[trim=0cm 3cm 0cm 0cm,clip=true,width=0.23\linewidth]{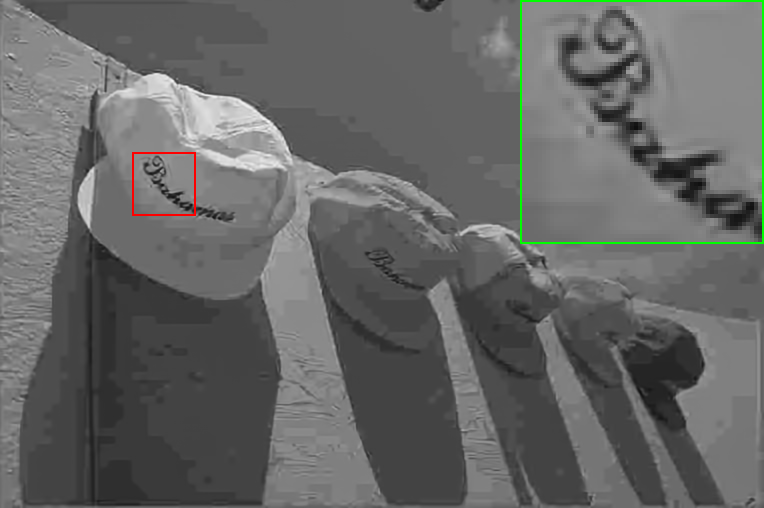}}

\subfigure[FBCNN (32.04dB)]{
\includegraphics[trim=0cm 3cm 0cm 0cm,clip=true,width=0.23\linewidth]{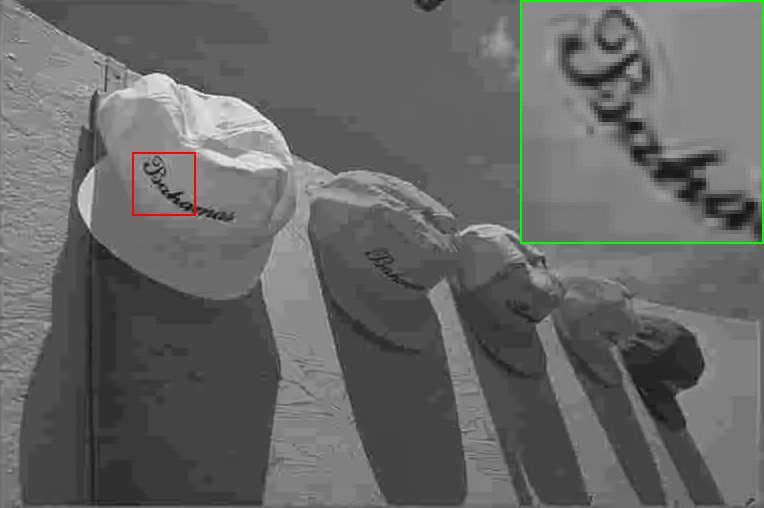}}
\subfigure[FBCNN-D (32.89dB)]{
\includegraphics[trim=0cm 3cm 0cm 0cm,clip=true,width=0.23\linewidth]{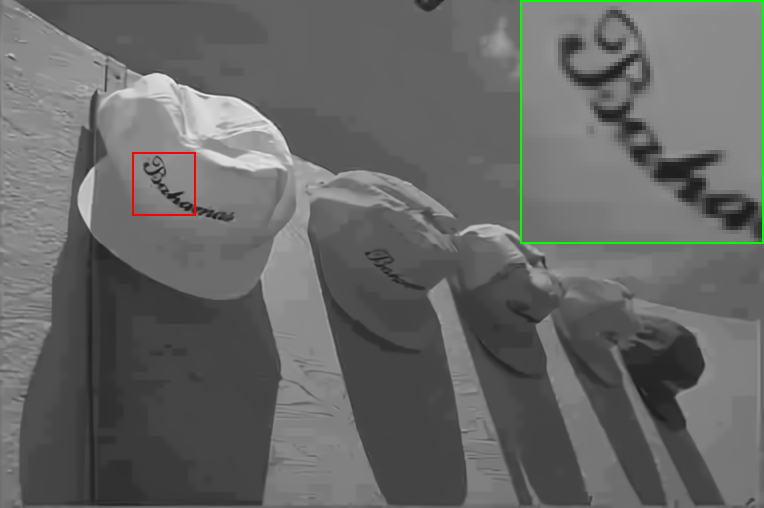}}
\subfigure[FBCNN-A (33.62dB)]{
\includegraphics[trim=0cm 3cm 0cm 0cm,clip=true,width=0.23\linewidth]{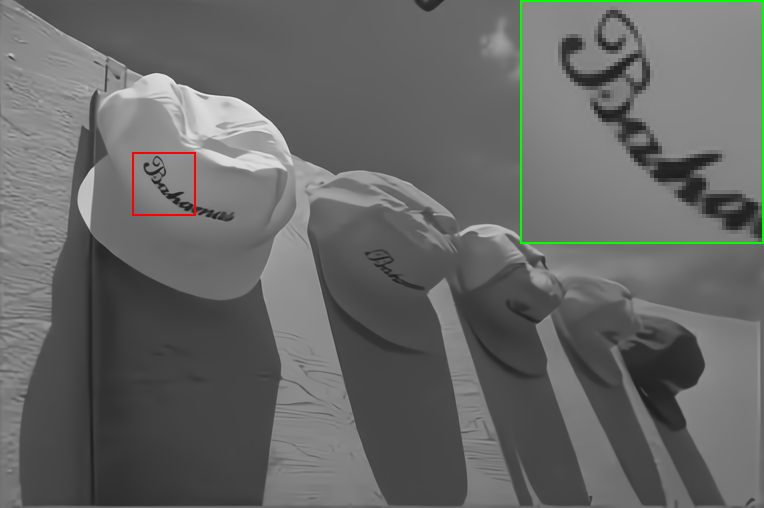}}
\subfigure[Ground Truth]{
\includegraphics[trim=0cm 3cm 0cm 0cm,clip=true,width=0.23\linewidth]{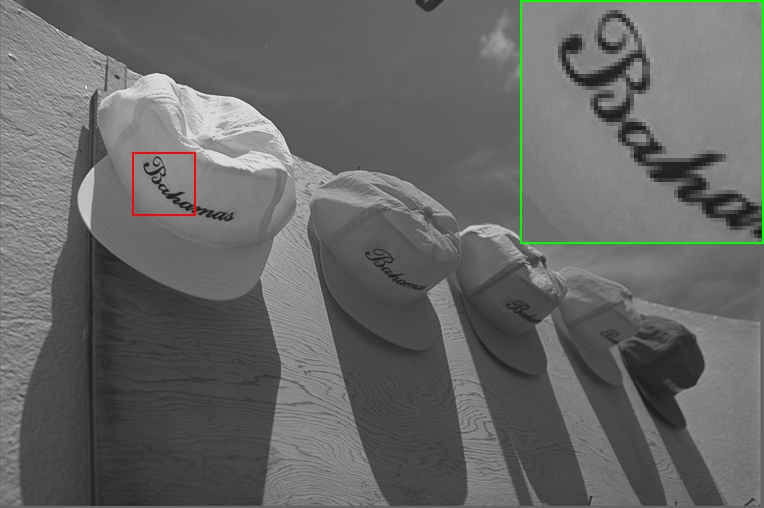}}
\caption{Visual comparisons of image `LIVE1: caps' with \textbf{non-aligned double} JPEG compression. This image is degraded by the first JPEG with $\text{QF}_1$ $=$10, pixel shift $=$ (4, 4), the second JPEG with $\text{QF}_2$ $=$ 30 successively.}
\label{doublejpegcompare}
\end{figure*}

\paragraph{Color JPEG image restoration} We also train our model on RGB channels, referred to as FBCNN-C. We compare FBCNN-C with QGAC, which is a state-of-the-art method, especially for color JPEG image restoration. The evaluation is made on LIVE1~\cite{sheikh2005live}, testset of BSDS500~\cite{martin2001database}, and ICB~\cite{icb} dataset. Although QGAC is specially designed for color JPEG image artifacts removal, we still get better performance by setting the input/output channels as 3. The result is shown in Table~\ref{tablecolor}.

\paragraph{Flexible JPEG image restoration} To demonstrate the flexibility of FBCNN, we show an example in Fig.~\ref{fig: flexibility}. By setting different quality factors, we can get results with different perception qualities. Users can make an interactive selection according to their preferences.

\subsection{Double JPEG Image Restoration}
The focus of our paper is to remove the complex double JPEG compression artifacts, which is an important step towards real image restoration. So we also evaluate the performance of current state-of-the-art methods and our proposed methods on images with double JPEG compression. We compare our methods with blind methods: DnCNN, DCSC, QGAC. The comparison is conducted using different combinations of quality factors ($\text{QF}_1$, $\text{QF}_2$) on the LIVE1 dataset. Each original image is JPEG compressed with $\text{QF}_1$, cropped by a random shift (4, 4) to the upper left corner, and JPEG compressed with $\text{QF}_2$.

\begin{figure*}[!ht]
\centering
\hspace{0.1cm}
\subfigure[JPEG]
{\includegraphics[width=0.126\textwidth]{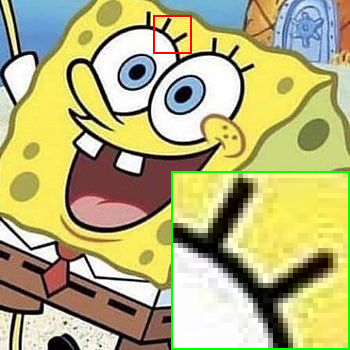}
}
\subfigure[DnCNN]
{\includegraphics[width=0.126\textwidth]{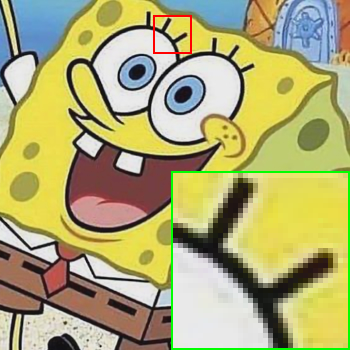}
}
\subfigure[DCSC]
{\includegraphics[width=0.126\textwidth]{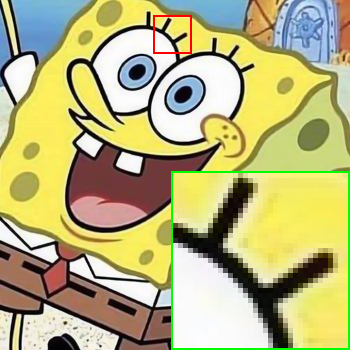}
}
\subfigure[QGAC]
{\includegraphics[width=0.126\textwidth]{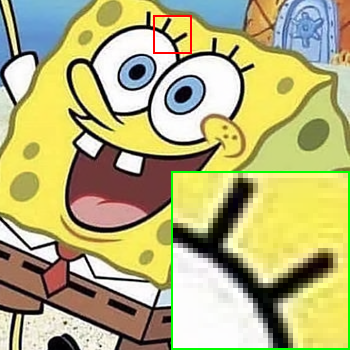}
}
\subfigure[FBCNN]
{\includegraphics[width=0.126\textwidth]{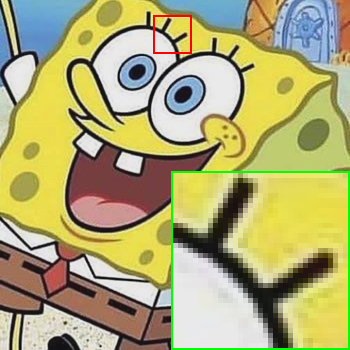}
}
\subfigure[FBCNN-D]
{\includegraphics[width=0.126\textwidth]{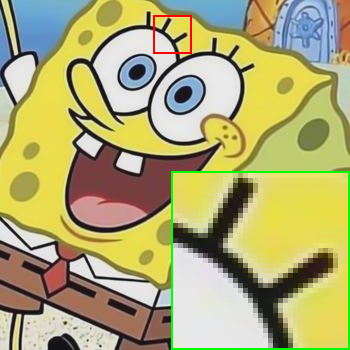}
}
\subfigure[FBCNN-A]
{\includegraphics[width=0.126\textwidth]{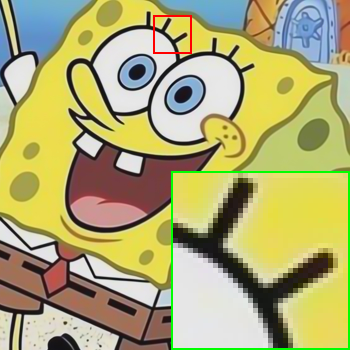}
}
\caption{Visual comparisons of an example from our Meme dataset.}
\label{real_compare}
\end{figure*}

The numerical and visual results are reported in Table~\ref{doublejpegtable} and Fig.~\ref{doublejpegcompare}. As shown in Table~\ref{doublejpegtable}, when changing the order of $\text{QF}_1$ and $\text{QF}_2$, although the differences between the PSNR values of JPEG images are generally smaller than 0.05 dB, a significant drop in performance can be seen on other methods and our FBCNN. Since DCSC is only trained with small quality factors from 10 to 40, it generally performs better than DnCNN, QGAC, and FBCNN when $\text{QF}_1$ $<$ $\text{QF}_2$. Despite some benefits for double JPEG compression, it should be pointed out that it is not reasonable to use a model trained with low quality factors to tackle all kinds of JPEG images. When dealing with relatively high-quality images, it tends to give more blurry results.

We also examine the effectiveness of our proposed two solutions to non-aligned double JPEG restoration. FBCNN-D is obtained based on FBCNN by correcting the quality factor by dominant QF estimation during inference. FBCNN-A is obtained by augmenting the training data with our proposed double JPEG degradation model. Table \ref{doublejpegtable} shows that by correcting the predicted quality factor, FBCNN-D largely improves the PSNR when $\text{QF}_1$ $<$ $\text{QF}_2$. FBCNN-A further improves performance when $\text{QF}_1$ $<$ $\text{QF}_2$. The difficult case when $\text{QF}_1$ $=$ $\text{QF}_2$ also sees an improvement on FBCNN-A.

\subsection{Real-World JPEG Image Restoration}

Besides the above experiments on synthetic test images, we also conduct experiments on real images to demonstrate the effectiveness of the proposed FBCNN. We collect 400 meme images from the Internet, as this kind of image is often compressed many times. Fig.~\ref{real_compare} shows a test example on our collected Meme dataset. 
Since there are no ground-truth high-quality images and no reliable no-reference image quality assessment (IQA) metrics, we do not report the quantitative results. We leave the study of no-reference IQA for JPEG compression artifacts removal for future works.

\section{Conclusions}
In this paper, we proposed a flexible blind JPEG artifacts removal network (FBCNN) for real JPEG image restoration. FBCNN decouples the quality factor from the input image via a decoupler and then embeds the predicted quality factor into the subsequent reconstructor through a quality factor attention block for flexible control. The predicted quality factor can also be adjusted to achieve a balance between artifacts removal and details preservation. Besides, we address non-aligned double JPEG restoration tasks to take steps towards real JPEG images with severe degradations. Extensive experiments on single JPEG images, the more general double JPEG images, and real-world JPEG images demonstrate the flexibility, effectiveness, and generalizability of our proposed FBCNN for restoring different kinds of degraded JPEG images.

\noindent\textbf{Acknowledgments:}
This work was partly supported by the ETH Z\"urich Fund (OK) and a Huawei Technologies Oy (Finland) project.

\clearpage

{\small
\bibliographystyle{ieee_fullname}
\bibliography{fbcnn}
}

\end{document}